\documentclass[twocolumn]{article}
\usepackage{amsmath}
\usepackage{latexsym}
\usepackage{graphicx}
\usepackage{amsfonts}
\usepackage{braket}
\usepackage{amssymb}
\usepackage{fancyhdr}
\usepackage[utf8]{inputenc}
\usepackage{dsfont}
\usepackage{colortbl}
\usepackage{xcolor}
\usepackage{subfigure}
\usepackage{psfrag}
\usepackage[colorlinks=true,citecolor=blue,linkcolor=blue]{hyperref}
\usepackage{tikz}
\usepackage{ulem}
\usepackage{bbold}
\usepackage{bbm}
\usepackage{titling}
\usepackage[top=3cm]{geometry}
\allowdisplaybreaks

\newcommand\bbone{\ensuremath{\mathbbm{1}}}
\usetikzlibrary{arrows,shapes}

\newcommand{\Trace}[1]{\text{Tr}\left[#1\right]}		

\newcommand{\ketbra}[2]{\ket{#1}\hspace*{-0.12cm}\bra{#2}}
\newcommand{\Ha}{\ensuremath{\text{H}}}
\newcommand{\norm}[1]{\left\lVert#1\right\rVert}
\newcommand{\qo}{{\ensuremath{\text{Q}}}}

\begin{document}

\title{Environmental effects in quantum annealing}
\author{Tobias Chasseur$^1$, Stefan Kehrein$^2$, and Frank K. Wilhelm$^1$}
\date{\vspace{-5ex}}
\maketitle

For quantum annealing, as opposed to circuit based quantum computing, the solution to a computational problem is encoded in the ground state of a quantum system. Therefore its susceptibility to environmental effects is a different but not less important open question essential to a scalable implementation. In this work we use renormalization group techniques to study the effect of Ohmic baths in a regime typical for practical adiabatic quantum computation. We show qualitative change to the effective Hamiltonian as well as the reduced qubit density matrix  encoding entanglement between system and bath. An effective dephasing of the reduced density matrix limits the extractable information from many qubit entangled groundstates. We find that the annealing process is no longer restricted to the qubits and discuss possible drawbacks or benefits of annealing in the combined system--bath states.
\footnotetext[1]{Theoretical Physics, Saarland University, 66123 Saarbr\"ucken, Germany}
\footnotetext[2]{Institute for Theoretical Physics, Georg-August-Universit\"at G\"ottingen, Friedrich-Hund-Platz 1, 37077 G\"ottingen, Germany}

\section*{Introduction}
Quantum computation and simulation has the potential to fundamentally outperform its classical counterparts due to superior scaling for specific tasks. The advantage in relevant problems only emerges on a medium to large sized quantum computer \cite{Preskill2018}. Two approaches to quantum computation show very different strategies to scaling and the treatment of environmental influence: While gate based quantum computing requires fidelities to meet hard thresholds and relies on error correction codes to increase fidelities, adiabatic quantum computing is conjectured to be more tolerant against coupling to the environment \cite{Albash2015} or to even benefit from it \cite{Passarelli2018,Kechedzhi2016}. As a result current quantum annealing architectures operate with a larger number of qubits but less shielding from the environment \cite{Gibney2017}.

The robustness against the environment is limited; it is a known result in strong coupling physics that high frequency bath modes correspond to structural changes of the effective Hamiltonian, e.g., the suppression of coupling in a two level system as shown using renormalization group theory \cite{Leggett1987,Zeldovich1972} as well as experimentally \cite{Lupascu2009}. 
With increasing bath coupling the two--level--system undergoes a dissipative phase transition towards a complete suppression of coupling between the qubit states -- this would mark a breaking point for adiabatic quantum computation. In this work we use poor man's scaling to derive an effective Hamiltonian and reduced density matrix to describe the qubits influenced by Ohmic baths. We go beyond  weak coupling which can be treated with perturbative master equations \cite{Deng2013,Albash2015} but is insufficient for describing a scaled quantum annealer and rather focus on the more suitable locally coherent but globally dephased (LCGD) regime. We discuss implications for annealing algorithms with regard to the desired ground state and extractable information as well as potential drawbacks or benefits from system--bath entangled states. 

\section*{Problem setting}
A physical device geared towards quantum computation is build around a set of two--level--systems, i.e. qubits which can within limitations interact with each other as well as be controlled and measured by an outside observer. While a circuit model architecture relies on long qubit lifetimes and high fidelity implementation of universal gate sets \cite{Preskill2018,Mohseni2017} the computational power of quantum annealing is encoded in the ground state of a widely adjustable Hamiltonian \cite{Albash2018}. As such it is thought to be more robust against imperfections leading to a scale up in system size at the cost of the afforementioned worse shielding against the environment \cite{Gibney2017}. To account for that we describe the system bath dynamic employing the multi--spin--Boson model, i.e., a number $n$ of qubits coupled to bath modes
\begin{align}
 \Ha=\Ha_{\rm q}+\sum_k\frac{\sigma_Z^{i_k}}{2}\lambda_k(a_k+a_k^\dagger)+\sum_k\omega_ka_k^\dagger a_k.\label{eqn:MSBM}
\end{align}
Here $\Ha_{\rm q}$ is the Hamiltonian solely acting on the qubit system; $k$ denotes the bath mode and $i_k\in\{1,2,\dots n\}$ the respective qubit it is coupled to. While the above Hamiltonian implies separated bath modes for each qubit we will treat the effects of a shared bath as well. Regardless thereof, an Ohmic distribution of bath modes $J_i(\omega)=2\alpha_i\omega\Theta(\omega_c-\omega)$ has been found suitable to model environmental effects in the regarded frequency regime and will be used throughout this work \cite{Bylander2011}.

Renormalization group techniques are tools to treat high dimensional quantum systems with varying energy scales. The main ansatz is to translate coupling to strongly off resonant terms into small changes to the Hamilton at a preferred frequency scale hence reducing the relevant Hilbert space. The generalization from the single spin case mentioned above to many qubits however poses a nontrivial challenge and is a major part of this work.

Poor man's scaling is a simple yet effective take on renormalization \cite{Kehrein2006}. As is common for renormalization techniques it translates a high energy part of the Hilbert space, i.e. excited bath modes with with frequencies above $\omega_0$ which is chosen much larger than the groundstate energy, into effective changes to the qubit Hamilton. While a single iteration can be identified as the Lamb--shift \cite{Bethe1947}, revaluation of $\omega_0$ with regard to the effective Hamiltonian allows for iterative application of poor man's scaling. We use this approach to derive the effective Hamiltonian as well as the reduced density matrix of its ground state for arbitrary qubit Hamiltonian depending on the strength of the system bath coupling.

\section*{Results}
We write the qubit Hamiltonian as a sum of Pauli strings, i.e., $\Ha_{\rm q}\equiv\sum \Delta_{\vec{s}} \sigma_{s_1}\otimes\sigma_{s_2}\dots\sigma_{s_n}$ and require a continuous $\omega_0\gg\Delta_{\vec{s}}$.
Leaving technical details to the supplementary material we find
\begin{align}
\Delta_{\vec{s}}(\omega_0)=\Delta_{\vec{s}}\left(\frac{\omega_0}{\omega_c}\right)^{c(\vec{s})}\label{eqn:delom}
\end{align}
in leading order of effects. The suppression of specific Pauli strings in the Hamiltonian via poor man's scaling is depended on a combined bath coupling constant $c(\vec{s})=\sum_{i\in\mathcal{M}} \alpha_i$ with $\mathcal{M}=\{i|[\sigma_{s_i},\sigma_Z]\neq0\}$. Consequently $\Delta_{\vec{s}}$ is an either constant or monotonically increasing function of $\omega_0$, i.e., decreasing the effective cutoff frequency $\omega_0$ reduces the contribution of the $\Ha_{\vec{s}}$ not commuting with the coupling to the bath. Note that the more general model of bath modes coupling to multiple qubits produces additional effects on $\Ha_{\rm q}$; however those are negligibly small in comparison with equation (\ref{eqn:delom}).

First focusing on individual Pauli strings  for $c(\vec{s})=1$ one sees that $\Delta_{\vec{s}}$ is linear in $\omega_0$ thus marking the transition between complete and incomplete suppression of $\Delta_{\vec{s}}$. For $\Ha_{\rm q}\propto\sigma_X^i$ we reproduce the single qubit case with the known phase transition at $\alpha=1$. It is however important to remember that the above covers special cases where the qubit Hamilton consists of a set of Pauli strings that sponsor equal $c(\vec{s})$. In more general settings the assumption $\omega_0\gg \norm{\Ha_{\rm q}}$ is validated for the ensemble of $\Delta_{\vec{s}}$, e.g., complete suppression requires all relevant $c(\vec{s})\geq1$. However despite a final $\omega_0\neq0$ Pauli strings with $c(\vec{s})\geq1$ are suppressed significantly stronger than those with weaker bath coupling representing more gradual transitions.

It is important to note that while the eigenvalues of the effective Hamiltonian are the energies of the system, the corresponding eigenstates $\ket {\Psi_l}$ do not describe full qubit states but their renormalized projection onto a Hilbertspace with lower cutoff frequency. This encodes entanglement between the qubit system and the eliminated bath modes. To account for that we derive effective qubit operators $\qo_{\rm eff}$ that describe the system as seen by accessible measurement operators \qo; or rather we derive the reduced density matrix such that $\braket{\qo_{\rm eff}}=\Trace{\rho_{r}\qo}$. In the supplementary material we show that the transition of the reduced density matrix coincide with that of the Hamiltonian described in equation (\ref{eqn:delom}). Notably one can rewrite the effect of the bath modes interacting with qubit $i$ to
\begin{align}
\rho\rightarrow\frac{1+\left(\omega_0/\omega_c\right)^{\alpha_i}}{2}~\rho~+\frac{1-\left(\omega_0/\omega_c\right)^{\alpha_i}}{2}~\sigma_Z^i\rho\sigma_Z^i\label{eqn:depha}
\end{align} 
which translates to a dephasing of that qubit with probability $1-\left(\omega_0/\omega_c\right)^{\alpha_i}$. This notation is equivalent to the operator--sum representation \cite{Nielsen2009}. As the effects on different qubits commute we identify the effects of poor man's scaling with simultaneous local dephasing of all qubits. However it is important to differentiate between the gradual dephasing associated with the coherence time $T_2$ and the fixed dephasing occurring here which poses a limit to the measurable phase correlation on any qubit of the system.  Note that the impure groundstate is not in conflict with the foundations of thermodynamics: other than there, each qubit adds another bath to the system, so a limit of infinitesimal Bath coupling is never reached.

Entanglement is an essential feature of quantum technologies necessary to allow supremacy in comparison to its classical counterparts; so does a ground state that is a single qubit product state not provide an advantage for quantum annealing over classical computation. The classification of multipartite entanglement is less straightforward than for just two parties reflecting the complexity of increasingly large quantum systems, however the GHZ state $\ket{\Psi}_{GHZ}=\frac{1}{\sqrt{2}}(\ket{00\dots0}+\ket{11\dots1})$ is considered to be a maximally entangled state \cite{Greenberger2007}. 
Assuming equal $\alpha_i=\alpha$ and the GHZ state being the entangled ground state of an effective Hamiltonian we find that the dephasing of \mbox{equation (\ref{eqn:depha})} which accompanies the poor man's scaling reduces the off diagonal elements of the density matrix and therefore the phase relation between the two product states by a factor $(\frac{\omega_0}{\omega_c})^ {n\alpha}$.  This depends to great extend on the ratio of initial and final cutoff frequency and therefore the specific system settings; for large enough $n$ it converges to a complete mixture of product states $\ket{00\dots0}$ and $\ket{11\dots1}$ which is neither coherent nor entangled. We suspect a general trend to suppress many qubit multipartite entanglement as it implies long Pauli strings in the density matrix.

\begin{figure}[t!]
 \centering
 \includegraphics[width=0.47\textwidth]{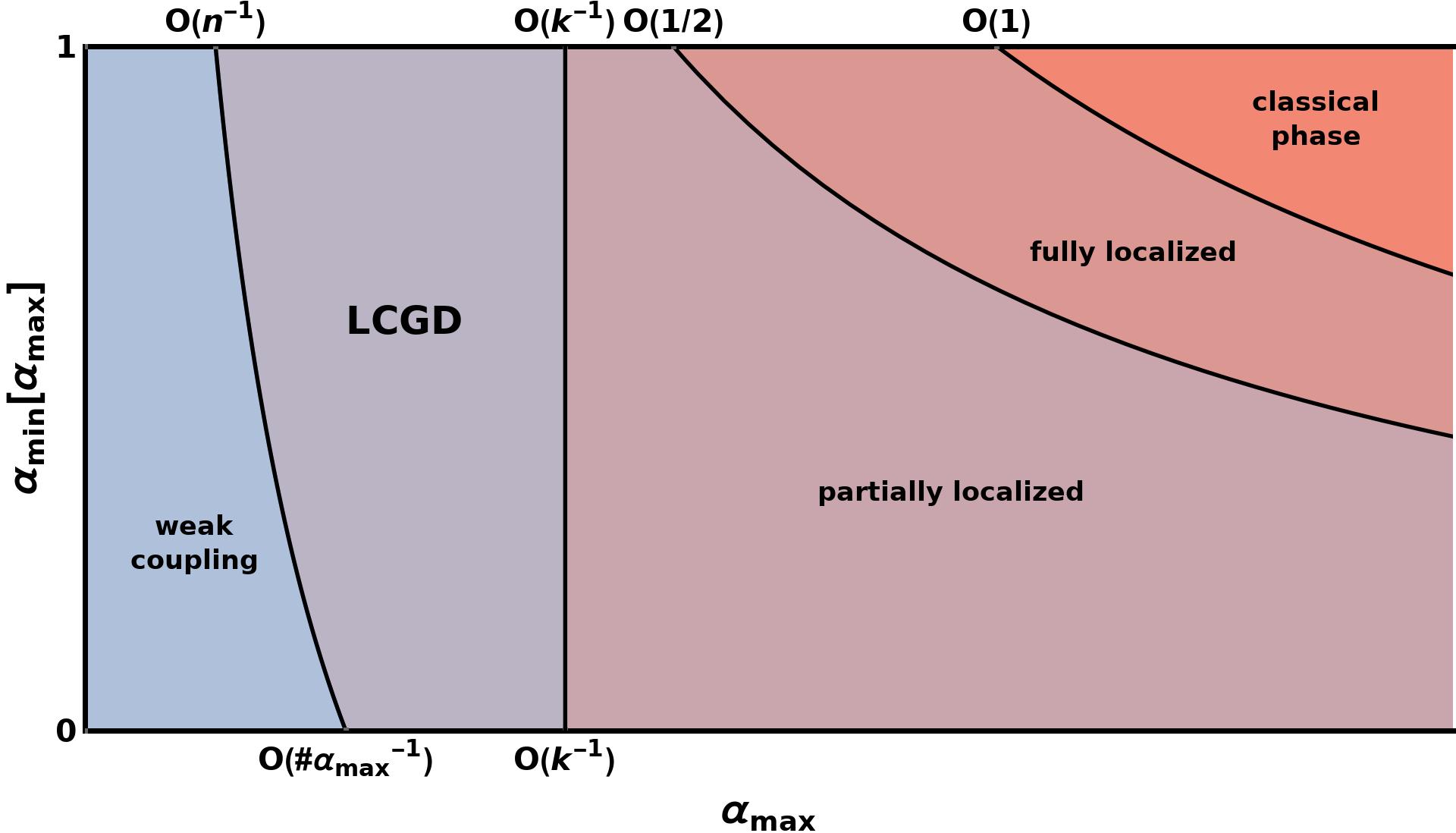}
 \caption{Diagram of different regimes of the $\nobreak{\alpha_i\in[\alpha_{min},\alpha_{max}]}$ for $n$ qubits. For simplicity we assume that all $\alpha_i$ are either minimal or maximal. We find that in the weak coupling regime all $c(\vec{s})$ are small, hence while the transformation changes the ratio between the different couplings it does not prohibit any Pauli strings. In contrast many--qubit $\sigma_{X/Y}$ strings are strongly suppressed in the LCGD regime while $k-$local interactions are only affected weakly. In the partially and fully localized regimes increasingly many $k-$local terms are suppressed limiting the possible ground states of the system. The classical phase describes a system where all terms commute hence the qubits' effective behaviour can be described classically.}
\label{fig:phase}
\end{figure}

\begin{figure}[t!]
 \centering
 \includegraphics[width=0.45\textwidth]{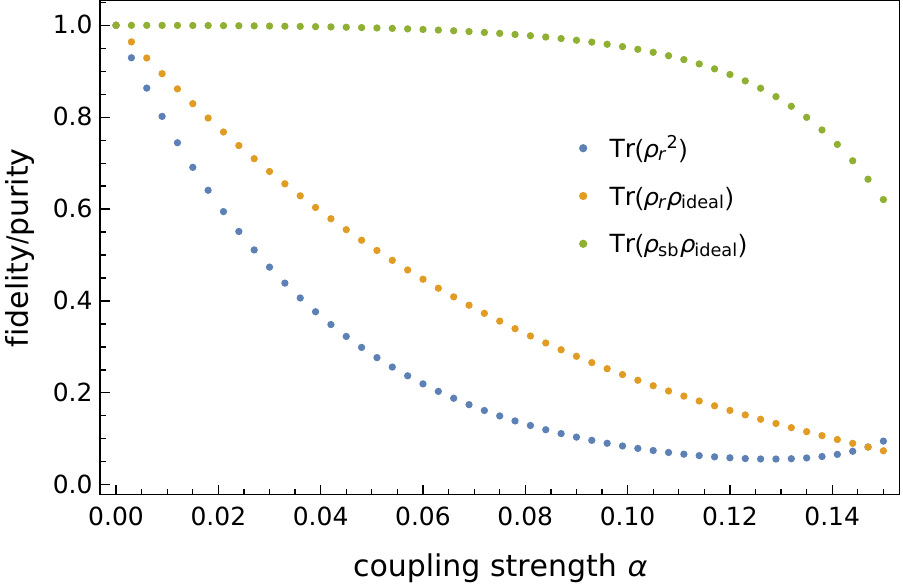}
  \caption{Analysis of the ground state of the antiferromagnetic Hamiltonian $\Ha_{\rm afm}$ for a specific $12$--qubit example at $s=0.8$. We calculated the ground state for both original and effective qubit Hamiltonian corresponding to minimal $\omega_0$  within $\Ha_{\rm eff}(\omega_0)\ll\omega_0$. The reduced density matrix is found by applying the transformation of equation (\ref{eqn:delom}) to the latter. One finds that while the system--bath state is relatively close to the intended ground state for $\alpha<0.1$ or $2\alpha<0.2$, the measurable state is not only considerably deformed but also incoherent at $\alpha=0.02$ which corresponds to $12\alpha=0.24$. Find additional figures including the entropy of the reduced density matrix in the supplementary material.}
 \label{fig:alpha}
\end{figure}

A Hamiltonian is $k-$local if all of its terms are acting nontrivially on at most $k$ qubits; those Hamiltonian can be used to engineer systems with arbitrary ground states for $k\geq2$ \cite{Kempe2006}. Such  a Hamiltonian is only effected by at most $k$ bath couplings which can be well controlled in experimental settings even when scaling to large qubit numbers. In contrast those pose a threat to the integrity of entangled density matrices as $\sum \alpha_i$ grows with the number of qubits. To further investigate that we are especially interested in the locally coherent but globally dephased (LCGD) regime where $k\alpha$ is small but $n\alpha$ is not. We showcase the regime in figure (\ref{fig:phase}). We focus on an antiferromagnetic Hamiltonian which is a realistic setting for quantum annealing application \cite{Hen2017}. We use the specific example
\begin{align}
\Ha_{\rm afm}&=s\sum_i\sigma_Z^i+s(1-s)\sum_{ij}c_{ij}\sigma_Z^i\sigma_Z^j\notag\\
&\quad+s\sum_{ij}c_{ij}\sigma_X^i\sigma_X^j
\end{align}
with $c_{ij}\in\{0,1\}$ only coupling specific qubit pairs and $s$ being the annealing parameter \footnote{courtesy of Itay Hen}. With an appropriate set of $c_{ij}\neq0$  this 2--local Hamiltonian can support multipartite entangled ground states. Showcasing a specific $12$--qubit example in figure (\ref{fig:alpha}) we verify the expected behaviour in the LCGD regime. While $2 \alpha$ is significantly smaller than one, the effective Hamiltonian is affected weakly resulting in a combined system bath density matrix $\rho_{\rm sb}$ close to the ideal state. Contrarily the qubit state as observed via external measurement is affected stronger indicating long Pauli strings in the density matrix. To show the prohibition of entanglement we computed the purity and entropy of $\rho_{r}$ as entanglement is limited by coherence. We find that in the regime where effective Hamilton and its groundstate are still intact the observable state is significantly to totally incoherent.

We see that in the LCGD regime, the effective Hamiltonian can sponsor a multipartite entangled state which however does not exist in the qubit space alone but also represent entanglement between qubit and high frequency bath modes; this needs to be taken into consideration when calculating system dynamics. Due to the effective dephasing of the reduced density matrix that ground state appears incoherent under qubit measurement.
\section*{Discussion}
We have investigated environmental effects represented by an Ohmic bath on quantum annealing using renormalization group theory. We derived the suppression of terms in the qubit Hamiltonian not commuting with the bath coupling which is increasingly strong for many--qubit terms. The same leading order transformation can be seen for the reduced density matrix encoding an entanglement between system and bath. We focus on the LCDG regime which inevitably arises when scaling to large systems, i.e. a robust $k$--local Hamiltonian can sponsor any multipartite entangled groundstate which is not accessible by qubit measurement. We discuss the implications for quantum annealing due to inaccessibility of the groundstate as well as the entanglement between system and bath.

The benefits of entanglement to annealing protocols often does not equal a many qubit entangled desired state. For example the adiabatic Grover or Deutsch-Josza algorithms do end up in product states \cite{Albash2015}, thus the readout of those is only perturbed weakly in the LCGD regime. However many--qubit entangled ground states occur e.g. in many quantum simulations where the final Hamiltonian represents a physical system that is hard to solve classically, i.e., is not a product state \cite{Babbush2014,Harris2018}. Those are limited in extractable information. More specifically the LCGD--dephased density matrix allows high fidelity measurements only for $k$--local operators while many--qubit correlations are not represented in the measurement. 

While the final Hamiltonian does not generally produce an entangled ground state, the system passes those on an efficient annealing path \cite{Lanting2014}. This means that while one starts and ends with a pure qubit state, the quantum annealing happens in both system and bath. Contrary to common intuition we believe that this does not per se prohibit quantum computation and may even be beneficial to the annealing process. We believe that the combined system--bath state no longer evolves adiabaticly but rather behaves more thermically, which has been shown to be beneficial \cite{Kechedzhi2016}. While experiments with low qubit numbers seem to support this reasoning \cite{Dickson2013} it remains to be investigated also with regards to its dependence on further architecture and protocol specifics as well as tested in a LCGD--regime device with $n\gg k$.

The multispin Bose model has been extensively studied (e.g. \cite{Schehr2006,Winter2014}) however mostly are not focused on spin dynamics which is required to investigate qubit based applications. Previous work studying environmental effects on quantum annealing however have been relying on perturbative master equations thus restricting themselves to weak coupling hence weak effects\cite{Deng2013, Albash2015}. With the renormalization group approach we go beyond this limit, hence are able to treat the LCGD regime between the weak and strong coupling limit investigated by Albash and Lidar \cite{Albash2015}. In that we could conclude that quantum annealing partially happens in both qubits and environment and only provides limited access to a multipartite entangled groundstate which however is sufficient for a lot of algorithms.

\section*{Acknowledgments}
We thank Itay Hen for providing the antiferromagnetic Hamiltonian. The research is based upon work (partially) supported by the Office of the Director of National Intelligence (ODNI), Intelligence Advanced Research Projects Activity (IARPA), via the U.S. Army Research Office contract W911NF-17-C-0050. The views and conclusions contained herein are those of the authors and should not be interpreted as necessarily representing the official policies or endorsements, either expressed or implied, of the ODNI, IARPA, or the U.S. Government. The U.S. Government is authorized to reproduce and distribute reprints for Governmental purposes notwithstanding any copyright annotation thereon.
\section*{Author contributions}
F.K.W. conceived the project. T.C derived the environmental effects together with S.K. and F.K.W. T.C. performed the numerical simulations under the supervision of F.K.W. The results were interpreted and the manuscript was written  by T.C. and F.K.W.
\newpage
\title{Environmental effects in quantum annealing:\\Supplementary Material}
\author{Tobias Chasseur$^1$, Stefan Kehrein$^2$, and Frank K. Wilhelm$^1$}
\date{\vspace{-5ex}}
\maketitle

\footnotetext[1]{Theoretical Physics, Saarland University, 66123 Saarbr\"ucken, Germany}
\footnotetext[2]{Institute for Theoretical Physics, Georg-August-Universit\"at G\"ottingen, Friedrich-Hund-Platz 1, 37077 G\"ottingen, Germany}
\section*{Transformation under poor man's scaling}
\emph{We derive the transition of effective Hamiltonian and reduced density matrix under poor man's scaling, which is a renormalization technique as described in Ref. \cite{sKehrein2006}.}

As is common for renormalization techniques poor man's scaling assumes a different energy scales within a system so that the Hilbert space can be seperated into a high energy and lower energy part, i.e, $\mathcal{H}\equiv\mathcal{H}_l\oplus\mathcal{H}_h$. There is no necessity for a distinct gap in the energy spectrum; just for the energy scales of $\mathcal{H}_h$ to be large in comparison to energies near the groundstate energy. The approach of poor man scaling is to derive an eigenwert equation on the lower energy subspace. For $\text{P}_{(l/h)}$ projecting onto the respective subspaces and $\text{H}_{(l/h)(l/h)^\prime}= \text{P}_{(l/h)}\text{H}\text{P}_{(l/h)^\prime}$ we find:
\begin{align}
 \begin{pmatrix}
  \Ha_{ll} && \Ha_{lh}\\
  \Ha_{hl} && \Ha_{hh}
 \end{pmatrix}
\begin{pmatrix}
  \ket{\Psi_l}\\
  \ket{\Psi_h}
 \end{pmatrix}
=E\begin{pmatrix}
  \ket{\Psi_l}\\
  \ket{\Psi_h}
 \end{pmatrix}\label{eqn:ansatz}
 \end{align}
 hence
 \begin{align}
\Ha_{hl}\ket{\Psi_l}+\Ha_{hh}\ket{\Psi_h}=E\ket{\Psi_h}\\
\ket{\Psi_h}=\left(E-\Ha_{hh}\right)^{-1}\Ha_{hl}\ket{\Psi_l}
\end{align}
and therefore
\begin{align}
\left(\Ha_{ll}+\Ha_{lh}\left(E-\Ha_{hh}\right)^{-1}\Ha_{hl}\right)\ket{\Psi_l}=E\ket{\Psi_l}.\label{eqn:nonlin}
\end{align}
Eq. \ref{eqn:nonlin} is exact but nonlinear and therefore not easily solvable for $E$. In the low energy regime $\Ha_{hh}$ is the dominant term in  $\left(E-\Ha_{hh}\right)$.  The standard ansatz is to approximate $E$ by the groundstate energy $E_{gs}$; however we found that this gives an unwanted shift in the effective change of the Hamiltonian resulting in non negligible errors. We therefore refrain from making the approximation while still assuming that $E$ is small compared to all energy levels 
of $\mathcal{H}_h$. This leads to a nonlinear equation for determining $E$ which can be linearized by approximations better suitable for the problem at hand. 

For that we look at a quantum system of multiple qubits with a Pauli $\sigma_Z$-coupling to a set of bath modes which we assume to be ohmicly distributed, i.e.:
\begin{align}
 \Ha=\Ha_q+\sum_k\frac{\sigma_Z^{i_k}}{2}\lambda_k(a_k+a_k^\dagger)+\sum_k\omega_ka_k^\dagger a_k
\end{align}
where $\Ha_q\equiv\sum \Delta_{\vec{s}} \sigma_{s_1}\otimes\sigma_{s_2}\dots\sigma_{s_n}$ acts solely on the qubit system. We define the Hilbert space separation \mbox{$\mathcal{H}_l=\mathcal{H}_q\otimes\mathcal{H}_{\omega_k<\omega_0}\otimes\{\ket{0}\}_{\omega_k\geq\omega_0}$} which excludes all state in which any bath mode above a $\omega_0$ is excited from the lower energy Hilbert space. This separation is not energy ordered, however it assures that any eigenenergy of $\Ha_{hh}$ is above $\omega_0$ which will be vital for the following. Firstly we look at
\begin{align}
\Ha_{lh}&=\text{P}_l\Ha_q\text{P}_h+\text{P}_l\sum_k\frac{\sigma_Z^{i_k}}{2}\lambda_k(a_k+a_k^\dagger)\text{P}_h\\&\quad+\text{P}_l\sum_k\omega_ka_k^\dagger a_k\text{P}_h\\
&=\text{P}_l\sum_k\frac{\sigma_Z^{i_k}}{2}\lambda_k(a_k+a_k^\dagger)\text{P}_h\\
&=\text{P}_l\sum_{k\geq k_0}\frac{\sigma_Z^{i_k}}{2}\lambda_ka_k\text{P}_h\\
\intertext{and}
\Ha_{hl}&=\text{P}_h\sum_{k^\prime\geq k_0}\frac{\sigma_Z^{i_{k^\prime}}}{2}\lambda_{k^\prime} a_{k^\prime}^\dagger\text{P}_l
\end{align}
We make the assumption that the only terms with $k=k^\prime$ contribute. This can be justified by neglecting second order terms of $\{a_k,a_k^\dagger\}$ in $(E-\Ha_{hh})^{-1}$ or, more rigorously, by applying the poor man scaling to every bath--mode individually. This restricts the inversion problem on the single excitation subspace of a specific mode , i.e., \mbox{$\mathcal{H}_h(k^\prime)=\mathcal{H}_q\otimes\mathcal{H}_{\omega_k<\omega_0}\otimes\{\ket{0}\}_{\omega_0\leq\omega_k\neq\omega_{k^\prime}}\otimes\{\ket{1}\}_{\omega_{k^\prime}}$}. In this subspace we find
\begin{align}
 \Ha_{hh}&=\Ha_q+\sum_{k<k_0}\frac{\sigma_Z^{i_k}}{2}\lambda_k(a_k+a_k^\dagger)+\sum_{k<k_0}\omega_ka_k^\dagger a_k\notag\\
 &\quad + \sum_{k\geq k_0}\frac{\sigma_Z^{i_k}}{2}\lambda_k(a_k+a_k^\dagger)+\sum_{k\geq k_0}\omega_ka_k^\dagger a_k\\
 &\overset{\mathcal{H}_h(k^\prime)}{=}\ketbra{1}{0}_{k^\prime}\Ha_{ll}\ketbra{0}{1}_{k^\prime}+0+\omega_{k^\prime}.
\end{align}
As we are focusing on the lowest energy eigenstates, $\omega_{k^\prime}$ is the dominant term. Approximating the inversion in Eq. \ref{eqn:nonlin} in first order of $\Ha_{ll}$ and $E$ on finds 
\begin{align}
(E&-\Ha_{ll}-\omega_{k^\prime})^{-1}\approx-\frac{1}{\omega_{k^\prime}}\left(1+\frac{E}{\omega_{k^\prime}}-\frac{1}{\omega_{k^\prime}}\Ha_{ll}\right)\label{eqn:inv}
\end{align}
and Eq. \ref{eqn:nonlin} transforming to 
\begin{align}
 E\ket{\Psi_l}&=\Ha_{ll}+\sum_{k^\prime\geq k_0}\frac{\lambda_k^{\prime}}{2}\sigma_Z^{i_k^\prime}\left[-\frac{1}{\omega_{k^\prime}}\left(1\right.\right.\notag\\
 &\quad \left.\left.+\frac{E}{\omega_{k^\prime}}-\frac{1}{\omega_{k^\prime}}\Ha_{ll}\right)\right]\frac{\lambda_k^{\prime}}{2}\sigma_Z^{i_k^\prime}\quad\ket{\Psi_l}\\
 &=\Ha_{ll}+\sum_{k^\prime\geq k_0}\frac{\lambda_{k^{\prime}}^2}{4\omega_{k^\prime}^2}\sigma_Z^{i_{k^\prime}}(\Ha_{ll}-E)\sigma_Z^{i_{k^\prime}}\quad\ket{\Psi_l}.\label{eqn:hami}
\end{align}
As $E$ is a number commuting with $\sigma_Z^{i_{k^\prime}}$ one finds the leading order effective Hamiltonian via recursive application as 
\begin{align}
 \Ha_{\rm eff}&=\Ha_{ll}+\sum_{k^\prime\geq k_0}\frac{\lambda_{k^{\prime}}^2}{4\omega_{k^\prime}^2}\left(\sigma_Z^{i_{k^\prime}}\Ha_{ll}\sigma_Z^{i_{k^\prime}}-\Ha_{ll}\right)\notag\\
 &=\Ha_{ll}+\sum_{k^\prime\geq k_0}\frac{\lambda_{k^{\prime}}^2}{4\omega_{k^\prime}^2}\left(\sigma_Z^{i_{k^\prime}}\Ha_{q}\sigma_Z^{i_{k^\prime}}-\Ha_{q}\right)\label{eqn:heff}
\end{align}
with omitting constant terms and using that all $\sigma_Z$ commute. One can see that whether $\Ha_q$ is affected by the transformation depends on if it commutes with the $\sigma_Z^i$. We rewrite
\begin{align}
\Ha_q\equiv\sum \Delta_{\vec{s}} \sigma_{s_1}\otimes\sigma_{s_2}\dots\sigma_{s_n}
\end{align}
as $\sigma_Z^i\sigma_{\vec{s}_i}\sigma_Z^i=\pm\sigma_{\vec{s}_i}$. This implies the transformation of the $\Delta_{\vec{s}}$ under poor man scaling
\begin{align}
 \Delta_{\vec{s}}\rightarrow\Delta_{\vec{s}}\left(1-\sum_{k^\prime\geq k_0}^{i_{k^\prime}\in\mathcal{M}}\frac{\lambda_{k^{\prime}}^2}{2\omega_{k^\prime}^2}\right)
 \end{align}
with $\mathcal{M}=\{i|[\sigma_{s_i},\sigma_Z]\neq0\}$, i.e., each Pauli string is affected by those bath modes where it does not commute with the $\sigma_Z$ of the respective qubit; affecting bath modes effectively reduce $\Delta_{\vec{s}}$.

It is important to note that poor man scaling generates an equation to determine eigenenergies  as well as the projection of the eigenvectors onto a lower dimension Hilbert space. Therefore associated transformation does neither preserve the orthogonality nor the normalization of the eigenvectors. Consequently it is neither unitary nor does it necessarily preserve the Hermitian properties of the qubit Hamiltonian. This effect can be seen in higher order terms. As the effective Hamiltonian is no longer a pure qubit Hamiltonian it is important to understand the transformation of measurement accessible qubit operators $\qo$. As those do not couple high and low energy Hilbert spaces one can employ the ansatz in Eq. \ref{eqn:ansatz} for any eigenvector to find
\begin{align}
 \braket{\qo}&=\bra{\Psi_l} \qo\ket{\Psi_l}+\bra{\Psi_h} \qo\ket{\Psi_h}\\
 &=\bra{\Psi_l} \qo\ket{\Psi_l}\notag\\
 &\quad+\bra{\Psi_l}\Ha_{lh}{(E-\Ha_{hh})^{-1}}^\dagger\notag\\
 &\quad\phantom{bra{\Psi_l}\Ha_{lh}}\qo(E-\Ha_{hh})^{-1}\Ha_{hl}\ket{\Psi_l}\\
 &= \bra{\Psi_l} \qo+\sum_{k\geq k_0}\frac{\lambda_k^2}{4\omega_k^2}\sigma_z^{i_k}\qo\sigma_z^{i_k}\ket{\Psi_l}+\mathcal{O}(\frac{1}{\omega_k^3}).
\end{align}
Notably the transformed operator one finds in the above equation acts on the projected eigenvectors which are no longer normalized. The easiest solution is to exploit $\braket{\bbone}=1$; we renormalize $\qo$ 
to 
\begin{align}
 \qo_{\rm eff}=\qo+\sum_{k^\prime\geq k_0}\frac{\lambda_{k^{\prime}}^2}{4\omega_{k^\prime}^2}\left(\sigma_Z^{i_{k^\prime}}\qo\sigma_Z^{i_{k^\prime}}-\qo\right),
\end{align}
which is the same transformation as for the effective Hamiltonian. While this is a remarkable coincidence we find that it is not universally true by looking at higher orders or bath modes coupling to multiple qubits. To describe the transformation of operators in a compact form we define the reduced density matrix $\rho_{r}$ that describes the state accessible by qubit measurements such that $\braket{\qo_{\rm eff}}=\text{Tr}[\rho_{r}\qo]$. As the density matrix undergoes the same transition we identify
\begin{align}
\rho\rightarrow \left(1-\frac{\lambda_{k^{\prime}}^2}{4\omega_{k^\prime}^2}\right)~\rho+\frac{\lambda_{k^{\prime}}\sigma_Z^{i_{k^\prime}}}{2\omega_{k^\prime}}~\rho~\frac{\lambda_{k^{\prime}}\sigma_Z^{i_{k^\prime}}}{2\omega_{k^\prime}}\label{eqn:opa}
\end{align}
as local dephasing of the qubit $i_{k^\prime}$ \cite{sNielsen2009}. 

The assumption of each bath mode coupling to just one qubit is a simple approach to incorporate the incoherent nature of the environment. While the qubit--qubit interaction via bath modes is in fact limited by the coherence of the bath, the introduction of additional coupling is a known effect \cite{sDevoret2004}. To account for both effects we incorporate less local bath modes that couple to multiple qubits with varying strength. One finds
\begin{align}
\Ha_{lh}&=\text{P}_l\sum_{k^\prime\geq k_0}\sum_i\frac{\sigma_Z^{i}}{2}\lambda_{k^\prime}^ia_k\quad\text{P}_h
\end{align}
and therefore 
\begin{align}
 \Ha_{\rm eff}&=\Ha_{ll}+\sum_{k^\prime\geq k_0}\sum_{i,j}\frac{\lambda_{k^{\prime}}^i\lambda_{k^{\prime}}^j}{4\omega_{k^\prime}}\left[\frac{1}{\omega_{k^\prime}}\left(\sigma_Z^{i}\Ha_{q}\sigma_Z^{j}\right.\right.\notag\\
 &\quad\quad\quad\left.\left.-\sigma_Z^{i}\sigma_Z^{j}\Ha_{q}\right)-\sigma_Z^{i}\sigma_Z^{j}\right].\label{eqn:xxyz}
\end{align}
While the $i=j$ effects were discussed in Eq. \ref{eqn:heff} the remainder is dominated by an introduced $\sigma_Z\sigma_Z$ coupling between qubits $i$ and $j$. Those effects do not interfere as the generated terms commute with the bath coupling. The remaining terms are weaker by a factor $||\Ha_q||/\omega_{k^\prime}\ll1$ in the regime where poor man scaling is applicable. We therefore find that for any system where $\lambda_i\lambda_j\ll\lambda_i^2$ does not hold, the bath effects on the Hamilton are dominated by $\sigma_Z\sigma_Z$ couplings thus either all other effects are negligible or the effective Hamiltonian is mainly consisting of $\sigma_Z\sigma_Z$ couplings. We therefore assume $\lambda_i\lambda_j\ll\lambda_i^2$ to hold.

Nonetheless we characterize the remaining terms $1/\omega_{k^\prime}(\sigma_Z^{i}\Ha_{q}\sigma_Z^{j}-\sigma_Z^{i}\sigma_Z^{j}\Ha_{q})$ qualitatively. We find that firstly they vanish for $\sigma_{s_j}$ commuting with $\sigma_Z$, secondly they do not change the bath couplings a term commutes with and thirdly for $\sigma_{s_i}\in\{\bbone,\sigma_Z\}$ and $\sigma_{s_j}\in\{\sigma_X,\sigma_Y\}$ they are antihermitian. Note again, that nonhermitian terms are just artefacts of cutting the state vector; they do not compromise the energy eigenvalues.

When including non  local bath modes we find the transformation of measured operators differs from the effective Hamilton. Most remarkably the newly found pure $\sigma^i_Z\sigma^j_Z$ terms do not occur. However the renormalization proves to be more difficult as $\braket{\bbone}$ depends on the qubit state. One finds leading order
\begin{align}
\bra{\Psi_l} &\qo_{\rm eff}\ket{\Psi_l}=\bra{\Psi_l}\qo+\sum_{k^\prime\geq k_0}\sum_{i,j}\frac{\lambda_{k^{\prime}}^i\lambda_{k^{\prime}}^j}{4\omega_{k^\prime}^2}\notag\\
&\left(\sigma_Z^{i}\qo\sigma_Z^j-\qo\bra{\Psi_l}\sigma_Z^i\sigma_z^j\ket{\Psi_l}\right)\ket{\Psi_l}.
\end{align}
which has to equal $\text{Tr}[\rho_{r}\qo]$ sponsoring the transition
\begin{align}
\rho\rightarrow\left(1-\frac{\lambda_{k^{\prime}}^i\lambda_{k^{\prime}}^j}{4\omega_{k^\prime}^2}\text{Tr}[\rho\sigma_Z^i\sigma_Z^j]\right)~\rho+\frac{\lambda_{k^{\prime}}^i\lambda_{k^{\prime}}^j}{4\omega_{k^\prime}^2}~\sigma_Z^i \rho\sigma_Z^j.
\end{align}
While this transition is nonlinear, it does preserve the trace of the density matrix and it describes the dephasing derived above for $i=j$. The smaller remaining terms only come into effect for either $\sigma_{s_i}, \sigma_{s_j}\in\{\sigma_Z,\bbone\}$ or $\sigma_{s_i}, \sigma_{s_j}\in\{\sigma_X,\sigma_Y\}$ as otherwise the contributions are canceled out by switching $i$ and $j$. As for the transition of the Hamilton operator we find that the set bath couplings a term commutes with does not change meaning the newly derived terms do not interact with the predominant dephasing.

\begin{figure}[htbp!]
 \centering
 \includegraphics[width=0.45\textwidth]{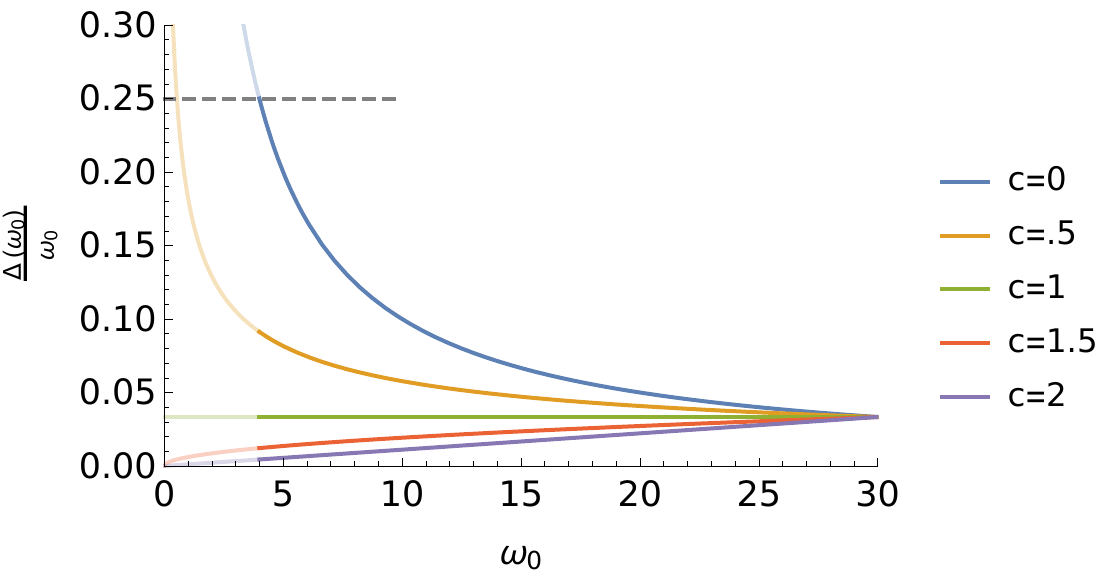}
 \caption{The ratio $\frac{\Delta_{\vec{s}}}{\omega_0}$ for different values of $c(\vec{s})$ starting at $\omega_0=30$ and $\Delta=1$. The indicated validity limit marks the abortion of poor man scaling as $\Delta_{c=0}\ll\omega_0$ no longer holds.}
 \label{fig:deom}
\end{figure}

To quantify the effects described above on a physical system the relevant bath modes can typically be described by an Ohmic distribution $J(\omega)=2\alpha\omega\Theta(\omega_c-\omega)$ with cutoff frequency $\omega_c$ and $\alpha$ mediating the coupling strength \cite{sBylander2011}. We focus on separated baths and make a transition from discrete to continuous modes, i.e. $\nobreak{\sum_k\lambda_k^2\rightarrow\sum_i\int J_i(\omega)\text{d}\omega}$. For an infinitesimal scaling $\omega_0=\omega_c-\delta\omega$ one finds the transition
\begin{align}
 \Delta_{\vec{s}}(\omega_c-\delta\omega)&=\Delta_{\vec{s}}(\omega_c)\left(1-\delta\omega\sum_{i}^{[\sigma_{\vec{s}},\sigma_Z^{i_{k^\prime}}]\neq0}\frac{\alpha_i}{\omega_c}\right)
 \intertext{and}
 \frac{\text{d} \Delta_{\vec{s}}}{\text{d}\omega_0}&=\frac{\Delta_{\vec{s}}}{\omega_0}\sum_{i}^{[\sigma_{\vec{s}},\sigma_Z^{i_{k^\prime}}]\neq0}\alpha_i\\
 &\equiv\frac{c(\vec{s})\Delta_{\vec{s}}}{\omega_0}\\
\Delta_{\vec{s}}(\omega_0)&=\Delta_{\vec{s}}(\omega_c)\left(\frac{\omega_0}{\omega_c}\right)^{c(\vec{s})}
\end{align}
with $0<c(\vec{s})<\sum \alpha_i$ a constant specific to the bath couplings a Pauli string does not commute with. We find $\Delta_{\vec{s}}$ to be either constant or monotonically increasing with $\omega_0$, i.e., decreasing the effective cutoff frequency $\omega_0$ reduces the contribution of the Pauli strings not commuting with all bath couplings. The above derivations are valid as long as the high energy frequencies are much larger than the effects of the qubit Hamilton hence poor man scaling can be applied until the condition is no longer met. As this implies $\Delta_{\vec{s}}\ll\omega_0$ we examine the dependencies of $\frac{\Delta_{\vec{s}}}{\omega_0}$ as shown in Fig \ref{fig:deom}. One can see that for
\begin{itemize}                                                                                                                                                                                                                                                                                                                                                                                                                                                                                                                                                              \item $c(\vec{s})=0$: Pauli strings that align with the bath coupling or act as the identity are not affected by the bath and therefore independent of $\omega_0$. Notably this applies to the bath induced $\sigma_Z\sigma_Z$ separating those effects.                                                                                                                                                                                                                                                                                                                                                                                                                                                                                                                                                               \item $0<c(\vec{s})=0<1$: $\Delta_{\vec{s}}$ decreases slower than $\omega_0$ hence at some point $\Delta_{\vec{s}}\ll\omega_0$ is no longer met yielding a nonzero effective $\Delta$.                                                                                                                                                                                                                                                                                                                                                                                                                                                                                                                                                            \item $c(\vec{s})=1$: the ratio $\frac{\Delta_{\vec{s}}}{\omega_0}$ is constant. This case marks the transition between complete and incomplete suppression of the qubit Hamiltonian.                                                                                                                                                                                                                                                                \item $1<c(\vec{s})$: $\Delta_{\vec{s}}$ decreases faster than $\omega_0$ and therefore can vanish completely.                                                                                                                                                                                                                                                                                                                                                                                                                                                                                                                                                                 \end{itemize}

It is however important to note that for a nontrivial Hamilton the condition $\Delta_{\vec{s}}\ll\omega_0$ is to be treated jointly for all Pauli strings.

\section*{The antiferromagnetic Hamiltonian -- supplementary Plots}
\begin{figure}[htbp!]
 \centering
 \includegraphics[width=0.45\textwidth]{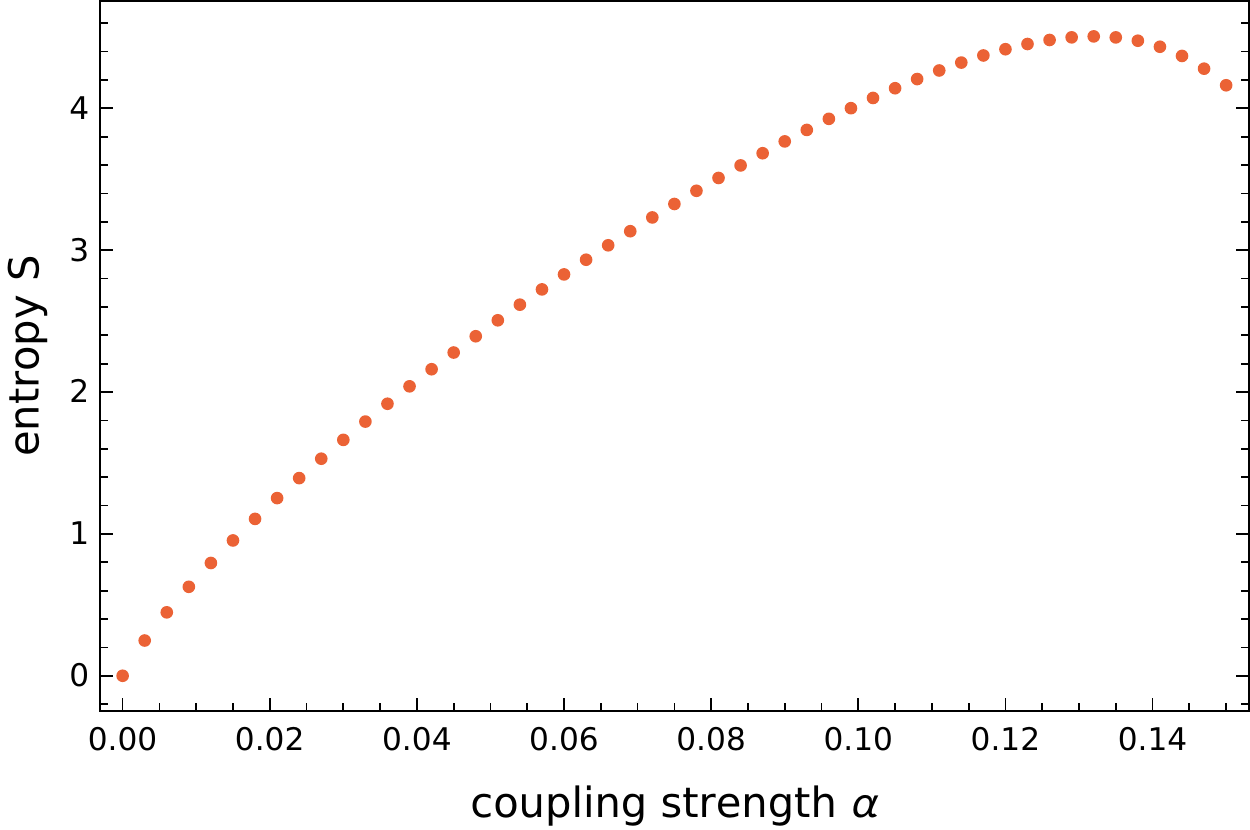}
 \caption{Entropy of the reduced density matrix of the $12$--qubit example used in the main work. Its increase even at small $\alpha$ indicates strong dephasing of a multipartite entangled ground state.}
 \label{fig:ea}
\end{figure}

\begin{figure}[t!]
 \centering
 \includegraphics[width=0.5\textwidth]{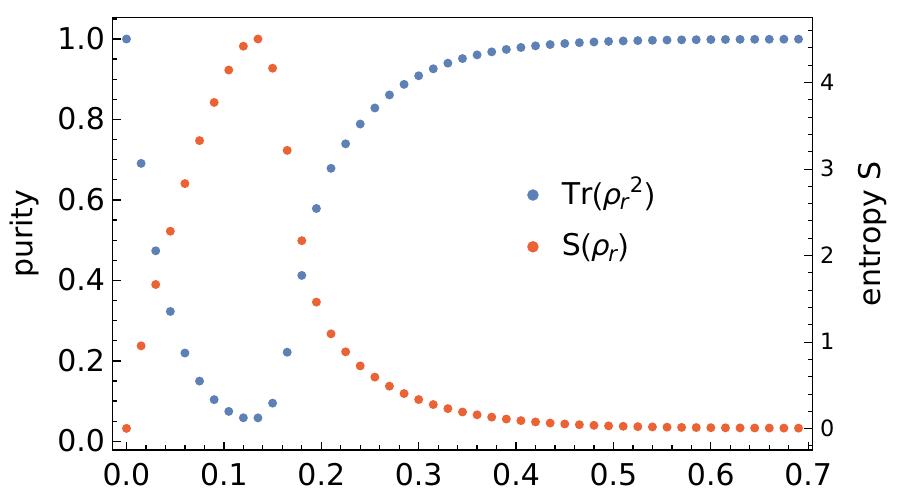}
 \caption{Purity and entropy of the reduced density matrix. For larger $\alpha$, $\rho_{\rm r}$ represents a pure state.}
 \label{fig:la}
\end{figure}

\begin{figure}[tbp!]
 \centering
 \includegraphics[width=0.45\textwidth]{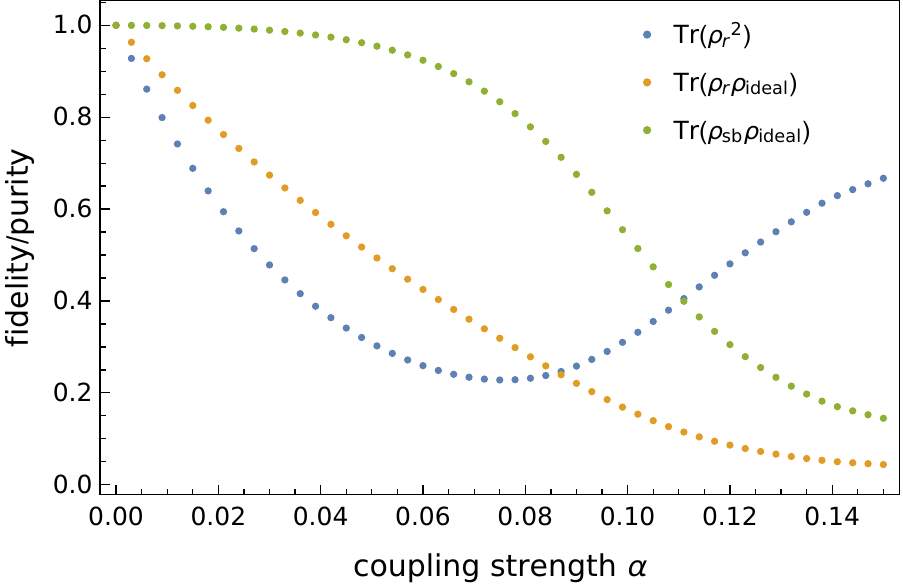}
 \caption{An instance of the $12$--qubit antiferromagnetic Hamiltonian generated by a different set of $c_{ij}$ at $s=0.7$. The deformation of the system--bath groundstate and revival of coherence take place for smaller $\alpha$.}
 \label{fig:na}
\end{figure}

We investigate the LCGD--regime, employing the antiferromagnetic Hamiltonian
\begin{align}
\Ha_{\rm afm}&=s\sum_i\sigma_Z^i+s(1-s)\sum_{ij}c_{ij}\sigma_Z^i\sigma_Z^j\notag\\
&\quad+s\sum_{ij}c_{ij}\sigma_X^i\sigma_X^j.
\end{align}
In addition to Fig. 2 of the main text we give additional plots showcasing the environmental effects on the Hamiltonian. In Fig. \ref{fig:ea} we show the entropy for the settings used before; the simulation confirms a significant loss of coherence even at $\alpha=0.02$. Notably, after a strong increase, the entropy declines again coinciding with the failure of the effective Hamiltonian. We see in Fig. \ref{fig:la} that in the limit of large $\alpha$ , i.e., strong effects on both Hamiltonian and reduced density matrix, the reduced density matrix represents a pure state. If all $\sigma_X\sigma_X$ terms in the Hamiltonian are eliminated, it solely consists of $\sigma_Z$ terms thus its eigenstates commute with those. As $\sigma_Z$ eigenstates are invariant under dephasing, the reduced density matrix is pure. However those states are product states hence the revival of coherence does not coincide with a revival of entanglement. In Fig. \ref{fig:na} we observe these effects for smaller $\alpha$ as the $\sigma_Z$ terms of the Hamiltonian are stronger for smaller annealing parameter $s$.   

\end{document}